# Soliton dynamics in the Ostrovsky equation with anomalous dispersion

R. Fariello, M.S. Soares, Y.A. Stepanyants

## 1. Introduction

The role of the Korteweg–de Vries (KdV) equation in the description of nonlinear wave phenomena in weakly dispersive media currently is well known (see, for example, (Ablowitz and Segur, 1981)). This equation successfully combines the competition of the main effects in the physics of nonlinear waves, dispersion and nonlinearity. After numerous works in the 1950s–1960s, which led to the derivation of the KdV equation in various media (fluids, plasma, solids, etc), the universal role of this equation was realized and formation of solitary waves (called solitons) from arbitrary initial perturbations was discovered through the numerical simulation firstly, and then, through the analytical study. Shortly afterwards, a wide class of completely integrable equations was discovered (see, for example, again (Ablowitz and Segur, 1981)).

In the meantime, it was found that there are equations of physical interest that are close to completely integrable ones but not solvable by developed analytical methods. The Ostrovsky equation which will be a matter of consideration in this paper is one of such examples. This equation was derived in 1978 (Ostrovsky, 1978) for the description of weakly nonlinear waves in the ocean but was later rederived for other types of waves; it has the following form:

$$\frac{\partial}{\partial x}\left(\frac{\partial \eta}{\partial t} + c\frac{\partial \eta}{\partial x} + \alpha\eta\frac{\partial \eta}{\partial x} + \beta\frac{\partial^3 \eta}{\partial x^3}\right) = \gamma\eta. \qquad (1.1)$$

Here $\eta$ denotes a perturbation in the medium (for example, a free surface in the case of water waves), $c$ is the speed of long linear waves, and $\alpha$, $\beta$, and $\gamma$ are coefficients that are determined by particular physical problems. The coefficient g is usually positive, whereas other two coefficients, $\alpha$ and $\beta$, can be of any sign.

For the past 40 years, the Ostrovsky equation has become popular in oceanography, plasma physics and many other fields. In particular, the Ostrovsky equation and its various versions describe wave processes in relaxing media (Vakhnenko, 1999), magnetised plasma (Obregon, Stepanyants, 1998) (even quark-gluon plasma (Fogaca et al., 2020)), and dielectrics (Kozlov, Sazonov, 1997). It also describes sound waves in a fluid with gas bubbles (Hunter, 1990), elastic waves in nonlinear chains (Yagi, Kawahara, 2001), ultrashort laser pulses (Litvak et al., 2005), and waves of any nature in randomly inhomogeneous media (Benilov, Pelinovskii, 1988) (see also the surveys (Grimshaw et al., 1998; Apel et al., 2007; Ostrovsky et al., 2015; Stepanyants, 2020)).

Considering waves of infinitesimal amplitudes, one can neglect the nonlinear term in Eq. (1) and search a simplest solution in the exponential form of $\eta \sim \exp[i(\omega t - kx)]$, where $\omega$ is a wave frequency and $k$ is a wavenumber. Substituting this solution into the linearized Ostrovsky equation (1), with $\alpha = 0$, we find the dispersion relation:

$$\omega = ck - \beta k^3 + \frac{\gamma}{k}. \tag{1.2}$$

From here we obtain the expressions for the phase and group velocities:

$$V_{ph} \equiv \frac{\omega}{k} = c - \beta k^2 + \frac{\gamma}{k^2}; \qquad V_g \equiv \frac{d\omega}{dk} = c - 3\beta k^2 - \frac{\gamma}{k^2}. \tag{1.3}$$

The dispersion relation $\omega(k)$ is shown in Fig. 1, and the dependences of phase and group velocities are given in Fig. 2.

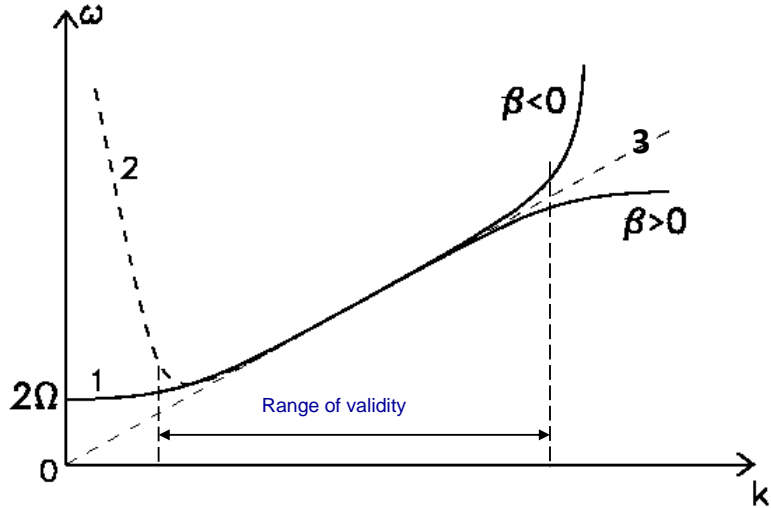

Fig. 1. Dispersion dependence of the wave frequency on the wave number for the Ostrovsky equation describing waves in a rotating ocean. Curve 1 corresponds to the full hydrodynamic equations of rotating fluid where $\Omega$ is the local Coriolis parameter (Grimshaw et al., 1998), dashed line 2 corresponds to Eq. 1.2. The case with $\beta > 0$ pertains to the normal dispersion, whereas the case with $\beta < 0$ pertains to anomalous dispersion. Dashed line 3 shows the dispersionless case.

As follows from Fig. 1, the Ostrovsky equation is applicable to wave processes of medium scales, where dispersion corrections to the straight dashed line 3 in the range of small and large wave numbers are relatively small. Thus, the condition of applicability of the Ostrovsky equation is reduced to the double inequality:

$$\sqrt{|\gamma|/c} \ll k \ll \sqrt{c/|\beta|}. \tag{1.4}$$

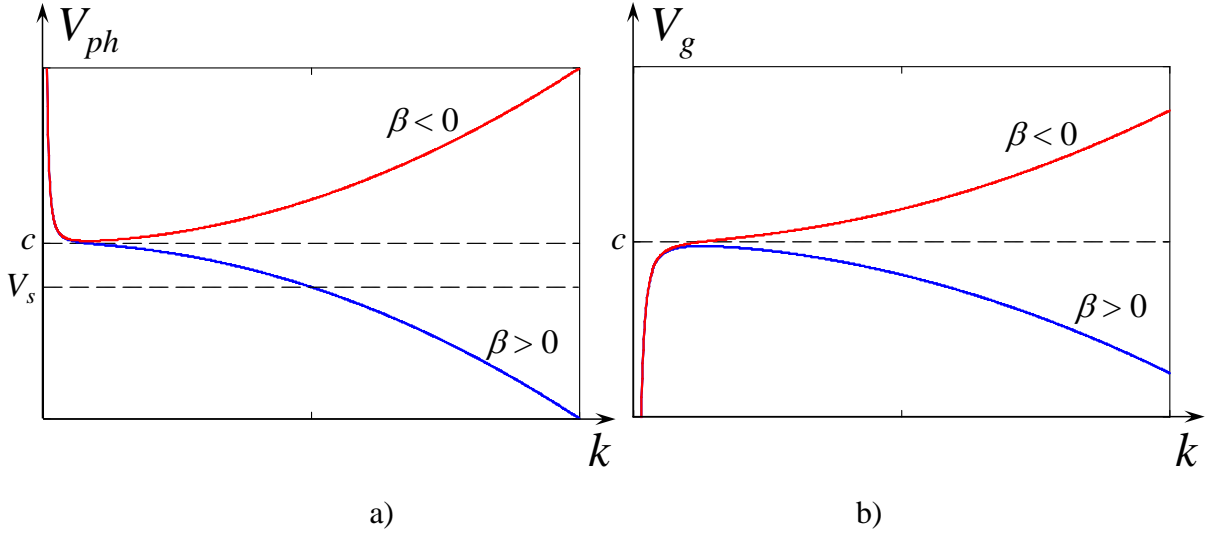

a) b)

Fig. 2. Qualitative dependences of the phase (left panel) and group (right panel) velocities on the wave number for different signs of the small-scale dispersion $\beta$. When $\beta > 0$, then a source moving with any speed $V_s$ is in the resonance with one of linear waves having the same phase speed. This usually leads to energy losses from the source to wave generation.

As one can see from Fig. 2a, when $\beta > 0$, the phase velocity of linear perturbations is unlimited, whereas the group velocity cannot exceed $c$ (see Fig. 2b). The peak group velocity is attained at $k_g = \sqrt[4]{\gamma/3|\beta|}$. The situation is reverse when $\beta < 0$, the phase velocity of linear perturbations is restricted from below, it cannot be less than $c$ (see Fig. 2a); whereas the group velocity is unlimited (see Fig. 2b). The minimum phase velocity occurs at $k_p = \sqrt[4]{\gamma/|\beta|}$. Such a dependence of the phase velocity on the wave number leads to the important physical consequence, stationary solitary waves can exist only when $\beta < 0$, whereas in the case of $\beta > 0$ their existence is impossible (Leonov, 1981; Galkin & Stepanyants, 1991).

The Ostrovsky equation complies with the following conservation laws:

- Mass conservation: $I_1 \equiv \int \eta(x,t)\,dx = 0$. (1.5)

- Energy conservation: $I_2 \equiv \frac{1}{2}\int \eta^2(x,t)\,dx = \text{const}$. (1.6)

- Hamiltonian: $I_3 \equiv \frac{1}{2}\int \left[\beta(\eta_x)^2 - \frac{\alpha}{3}\eta^3 - \gamma\upsilon^2\right]dx = \text{const.}, \quad \eta = \upsilon_x$. (1.7)

In these formulas, integration is assumed along the whole axis $x$ for localized perturbations or over the period for periodic perturbations. The constants in Eqs. (1.6) and (1.7), are determined by the initial conditions, whereas the mass conservation law is essentially the limitation on the class of admissible perturbations, which requires a zero mean value of perturbation per period or zero total mass of the localized perturbation. This limitation naturally follows from the condition of applicability of the Ostrovsky equation.

The case of the "anomalous dispersion", $\beta < 0$, $\gamma > 0$, when stationary solitary waves can exist, occurs in particular, for waves in magnetized rotating plasma (Obregon, Stepanyants, 1998), and perhaps for internal waves in shear flows (Alias et al., 2014). Solutions of the Ostrovsky equation in the form of solitary waves with the zero total mass dubbed further *Ostrovsky solitons* were numerically studied in (Obregon, Stepanyants, 1998). It was shown that their structure depends on the speed which is related to the soliton amplitude. The numerically obtained dependence of soliton amplitude on speed is shown in Fig. 3.

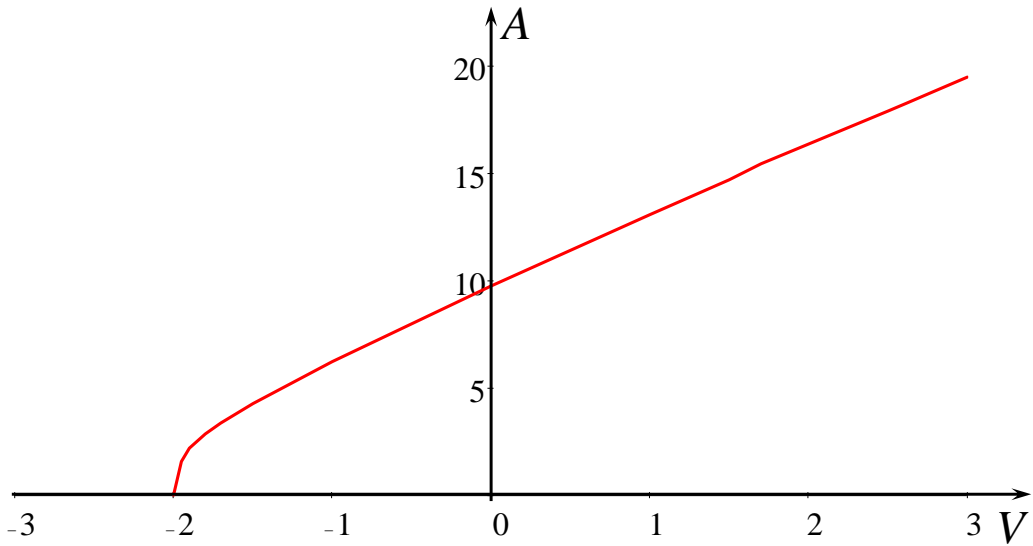

Fig. 3. The numerically obtained dependence of soliton amplitude $A$ on speed $V$ within the dimensionless Ostrovsky equation (2.3) (see below).

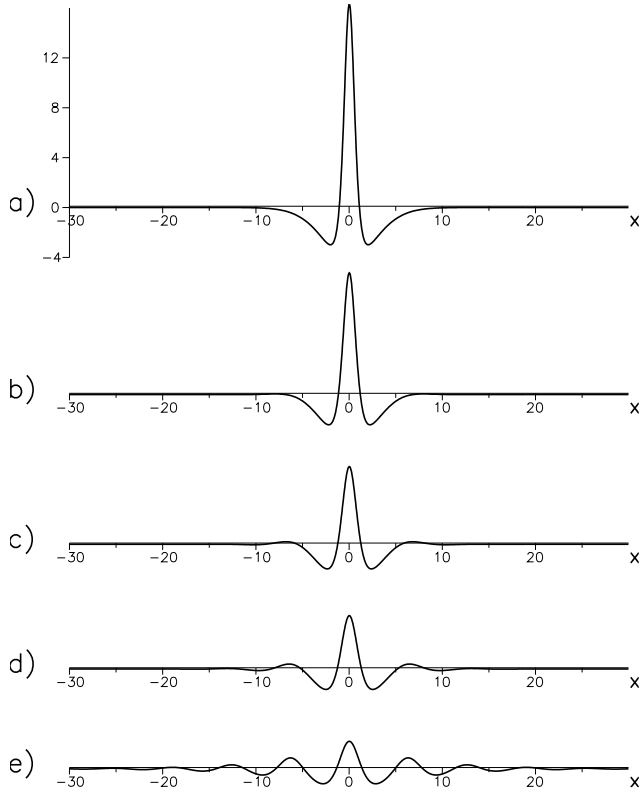 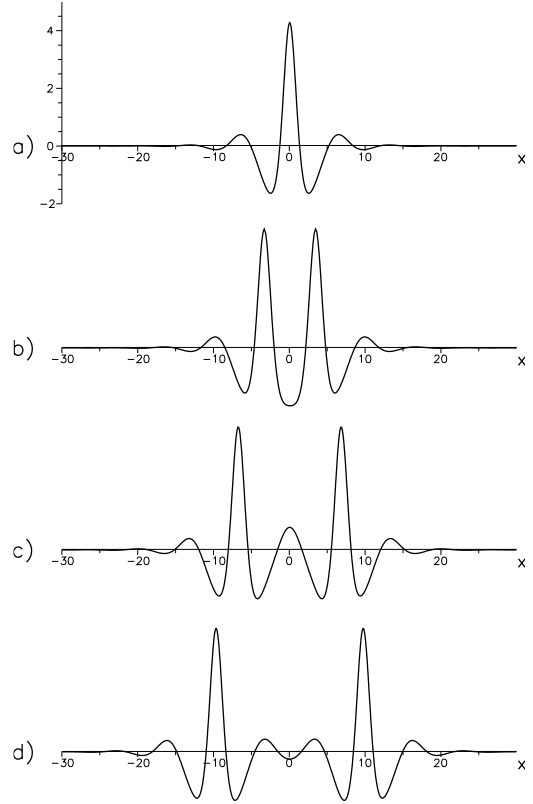

Fig. 4 (left). Variation of shapes of solitary waves with the amplitude in the Ostrovsky equation with anomalous dispersion at $\beta\gamma < 0$. From (Obregon, Stepanyants, 1998).

Fig. 5 (right). Examples of bound stationary states of two solitary waves in the case of anomalous dispersion $\beta\gamma < 0$. A single soliton with a nonmonotonic profile is shown in Fig. 5a, and on the subsequent figures Fig. 5b–Fig. 5d one can see a family of stationary bisolitons. From (Obregon, Stepanyants, 1998).

Typical soliton profiles are presented in Fig. 4; the total their masses are zero. As was shown in (Obregon, Stepanyants, 1998), solitons with small amplitudes, for example, like those presented in Fig. 4e, can be described by the nonlinear Schrödinger equation. The nonmonotonic character of a solitary wave profile implies that solitons can form stationary bound states: bisolitons, trisolitons, etc.; the examples of such stationary formations are shown in Fig. 5. Similar solutions were numerically constructed in (Whitfield & Johnson, 2016) and (Johnson, 2019). The bound states can be stable when a peak of each solitary wave coincides with a local minimum of the neighbouring soliton (see Figs. 5b, 5d) or unstable when the wave peaks coincide with local maxima of the neighbouring soliton (see Fig. 5c). Apparently, it is possible to create even an infinite stationary wave consisting of Ostrovsky solitons and multisolitons with random alternation of maxima and minima. Perhaps, two coupled solitons with nonmonotonic profiles can form an oscillating pair, termed a *breather*, similar to that observed in experiments with electromagnetic transmission lines; such breathers can emerge near stable states (Gorshkov et al., 1976; 2010). In periodic systems containing three and more bound solitons on each period, much more complex and even stochastic formations can arise (see, for example, (Gorshkov et al., 1976; 2010)).

Despite the existence of stationary solitary waves in the case of the "anomalous dispersion" when $\beta\gamma < 0$, their robustness and evolutionary character, i.e. possibility to emergence from arbitrary initial perturbations was not studied thus far. However, the adiabatic decay of an Ostrovsky soliton due to the cylindrical divergence was studied in (Fraunie & Stepanyants, 2002). That paper has demonstrated that the soliton keeps its structure in the process of adiabatic decay under the influence of small perturbation caused by the cylindrical divergence. The aim of this paper is to investigate the robustness of solitons in the Ostrovsky equation, possibility of their formation from arbitrary initial perturbations of a zero mass, their stability with respect to interaction with each other, and decay under the influence of dissipation.

## 2. The dimensionless form of the Ostrovsky equation and its soliton solutions

In this paper, we focus only on the case of the Ostrovsky equation with the "anomalous dispersion" assuming that $\beta\gamma < 0$. Then, Eq. (1.1) can be presented in the dimensionless form:

$$\frac{\partial}{\partial \xi}\left(\frac{\partial u}{\partial \tau} - u\frac{\partial u}{\partial \xi} - \frac{\partial^3 u}{\partial \xi^3}\right) = u \qquad (2.1)$$

with the help of the transformation:

$$\tau = -t\beta\left(-\frac{\beta}{\gamma}\right)^{3/4}, \quad \xi = (x - ct)\beta\left(-\frac{\beta}{\gamma}\right)^{1/4} \quad u = \frac{\alpha}{\beta}\left(-\frac{\beta}{\gamma}\right)^{1/2} \eta. \qquad (2.2)$$

Searching for a stationary solution, we assume that function $u$ depends on one variable $z = \xi - V\tau$, where $V$ is a dimensionless wave speed. Then, Eq. (2.1) reduces to the ODE:

$$\frac{d^2}{dz^2}\left(\frac{d^2 u}{dz^2} + Vu + \frac{1}{2}u^2\right) = -u. \qquad (2.3)$$

Analytical solutions of this equation are unknown. However, soliton solutions can be constructed numerically, for example, with the help of the Petviashvili method (for details see (Pelinovsky & Stepanyants, 2002)). As was aforementioned, solitary wave solutions depend on the wave speed $V$ which determines soliton amplitude. Soliton structure has been studied analytically and numerically in (Obregon, Stepanyants, 1998), and it was shown that for large negative values of $V < -2$, solitons have aperiodic exponential tails as shown in Fig. 4a and Fig. 4b. Within the range $-2 < V < 2$, solitons have oscillatory decaying asymptotics ("tails") as shown in Fig. 4c and Fig. 4d. The closer the parameter $V$ is to the upper critical value, $V_c = 2$, the more oscillations appear in the soliton shape, and solitons begin to look more and more like wavetrains (see Fig. 4e). However, it is important to note that for such solitons, carrier and envelope waves have the same velocities, in contrast to the usual envelope solitons of a general type within the nonlinear Schrödinger equation (Ablowitz and Segur, 1981). For $V > 2$, soliton solutions with zero asymptotics at infinity do not exist. Nevertheless, in this case other kinds of solitons matched with quasi-sinusoidal wavetrains can exist (see, e.g., (Grimshaw & Joshi, 1995)). Below we present a numerical study of soliton emergence from arbitrary initial pulses with zero masses.

## 3. Emergence of solitons from arbitrary initial pulses with zero masses

As a first step, let us check the emergence of Ostrovsky solitons from a quasi-soliton pulse. To this end, using the Petviashvili method, we constructed a soliton with aperiodic tails for $V = 2.615$. The stationarity of this solution was verified by another numerical code described in (Obregon, Stepanyants, 2012). It was verified that the solution is indeed stationary and stable, at least with respect to numerical rounding errors. Then, we used the same solution as the initial condition but multiplied it by the factor $F = 1.2$. During the evolution, the initial pulse emitted small-amplitude ripples and gradually slightly decreased. In a relatively short time, it quickly evolved into another stationary soliton with a bit smaller amplitude (see two upper frames in Fig. 6). A similar process was observed when the initial stationary soliton was multiplied by the factor $F = 0.9$ (see two bottom frames in Fig. 6). The dependences of pulse amplitudes on time in the process of soliton formation are shown in Fig. 7.

When the amplitude factor $F$ was small enough, the initial pulse transformed during propagation into another type of Ostrovsky soliton with a small amplitude and decaying oscillatory tails. Figure 8 illustrates an example of such an evolution when the amplitude factor is $F = 0.3$ (blue line in the top panel). The pulse amplitude decreased with time but then stabilized, approaching a constant level at $A = 0.58$ as in Fig. 9. As follows from these figures, solitons of a different shape with decaying oscillatory tails can also emerge from the initial pulse of small amplitude with aperiodic tails. This demonstrates that Ostrovsky solitons with both oscillatory and aperiodic tails are robust and can emerge from initial pulses in the process of evolution.

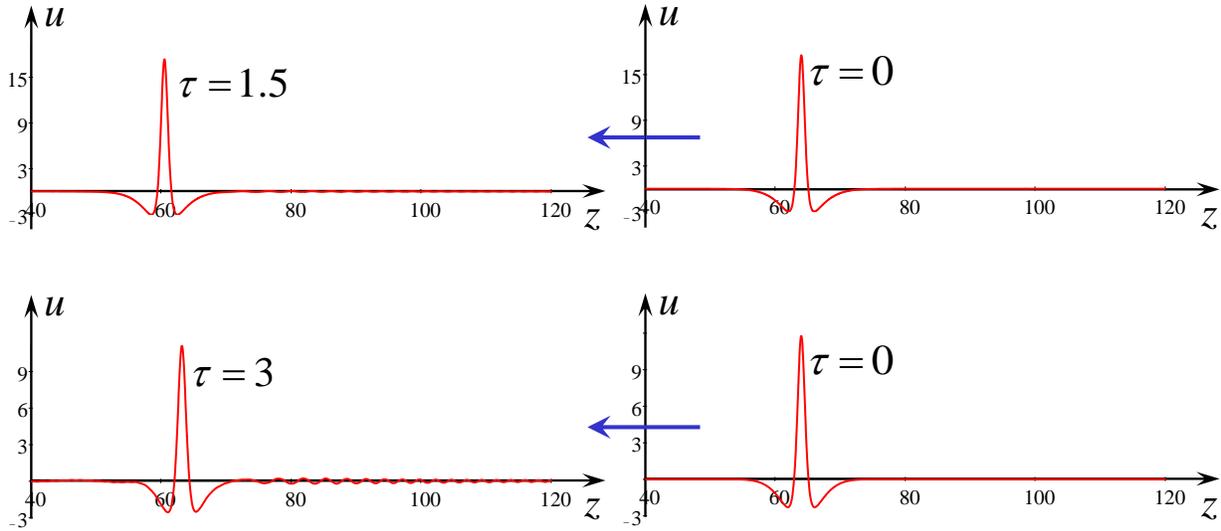

Fig. 6. Evolution of soliton-like initial pulses with a different amplitude factors $F$: in the top frames $F = 1.2$, in the bottom frames $F = 0.8$. Solitons move from right to left.

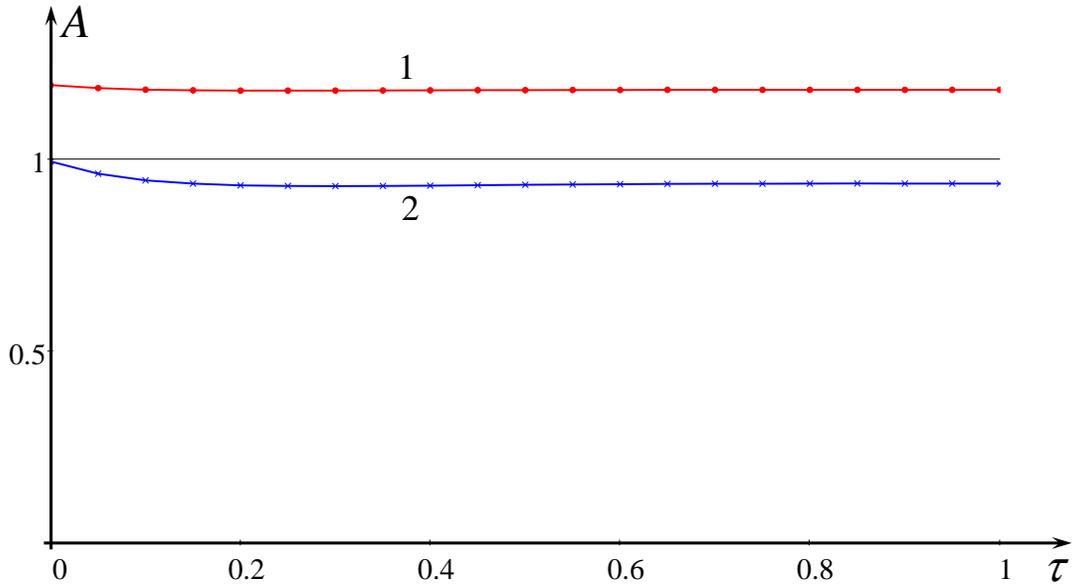

Fig. 7. Dependences of pulse amplitudes in the process of soliton formations. Line 1 pertains to the case with $F = 1.2$, and line 2 pertains to the case with $F = 0.8$.

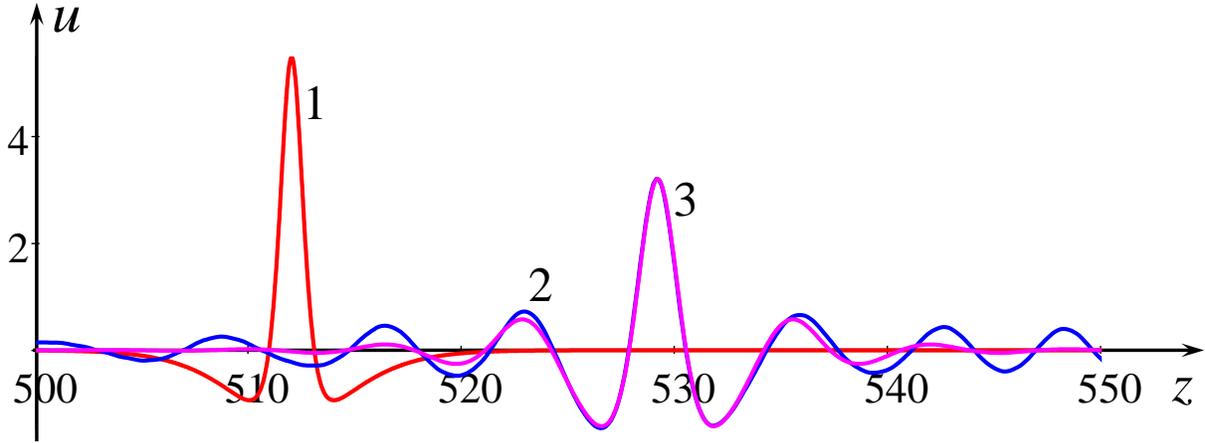

Fig. 8. The result of pulse evolution from $\tau = 0$ (red line 1) to $\tau = 10$ (blue line 2). Pink line 3 overlapping with the blue line 2 shows a stationary soliton with the same amplitude as the resultant pulse. (The authors are indebted to V.A. Gordin and D.P. Milutin for the data provided to this and the next figure 9.)

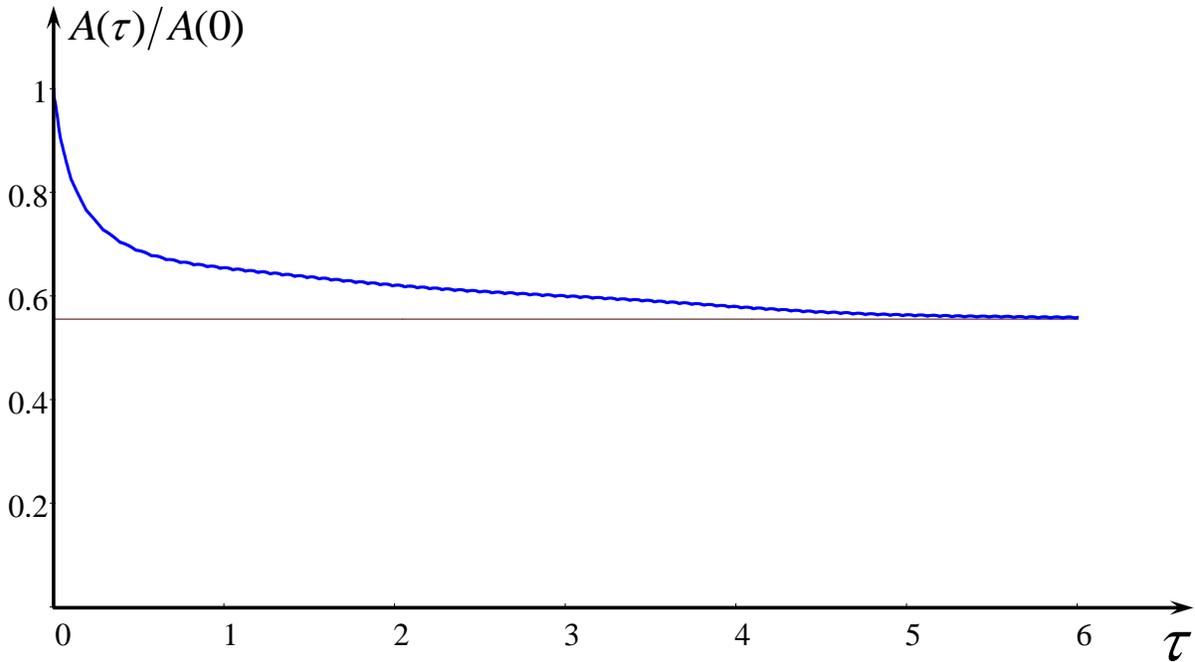

Fig. 9. Evolution of the amplitude of a soliton-like initial pulse with the amplitude factor $F = 0.3$. Horizontal line shows the asymptotic of the normalised soliton amplitude 0.555.

As the next step, we considered the decay of an initial pulse of notably larger amplitude into a number of solitons. For the initial condition, we again used a soliton with $V = 2.615$ and multiplied it by the factor $F = 10$. Then, the pulse disintegration into three solitons was observed; this is illustrated in Fig. 10. At the initial stage of evolution, the amplitude of the first pulse increased but then gradually decreased to a constant value, indicating that a stationary soliton had formed; this is illustrated in Fig. 11. This experiment demonstrates that solitons can emerge from intense initial pulses of the proper polarity, as in the KdV case.

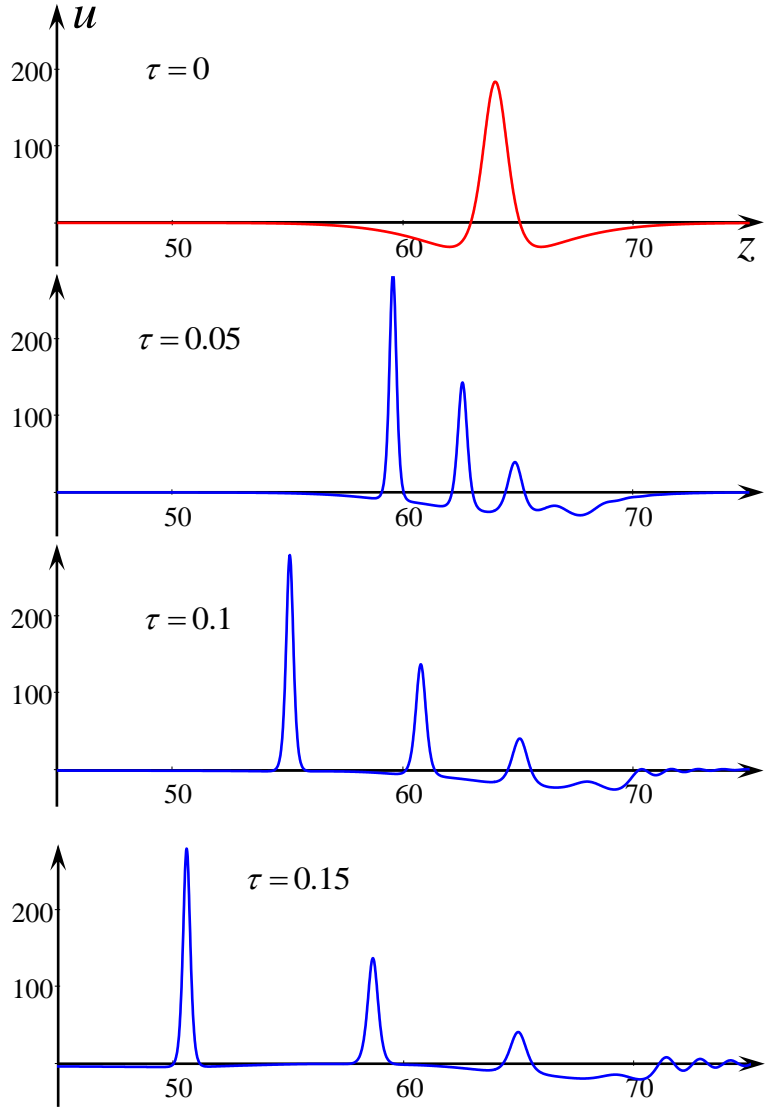

Fig. 10. Decay of the initial soliton-like pulse onto three Ostrovsky solitons of different amplitudes. Red line in the top panel shows the initial pulse, a soliton with the enhanced amplitude by the factor $F = 10$; blue lines in the subsequent panels from top to bottom show the pulse disintegration into three solitons and a quasi-linear wavetrain at $\tau = 0.05$, $\tau = 0.1$ and $\tau = 0.15$.

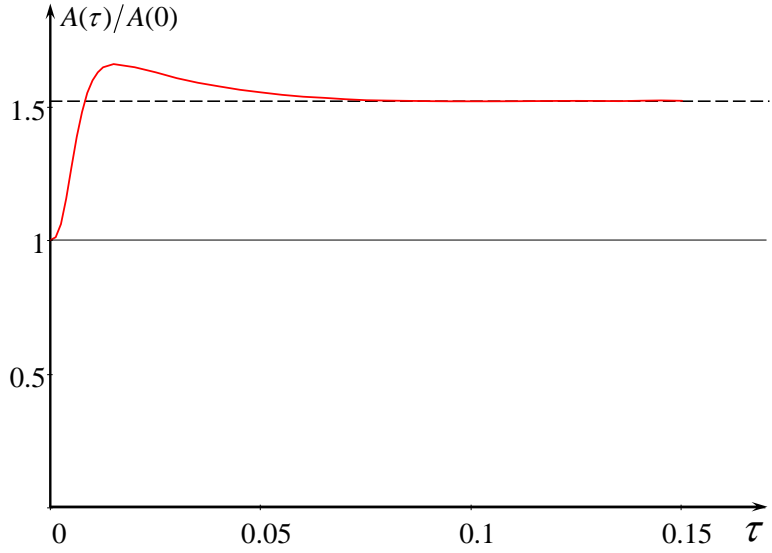

Fig. 11. Variation of the leading pulse amplitude in the process of soliton formation.

The dynamics of solitons becomes more complex when the pulse-type initial perturbation is wide and produces many solitons. In particular, we considered the initial perturbation in the form:

$$u(z,0) = A\frac{1-(z/D)^2}{\left[1+(z/D)^2\right]^2}, \qquad (3.1)$$

where $A = 50$ was the amplitude of the pulse and $D = 3$ was its characteristic width (see Fig. 12). The total mass of the initial pulse was $I_1 = 0$ as required by the Ostrovsky equation (see Eq. (1.5)). During evolution, the pulse initially disintegrated into 5 solitons and a trailing ripple, as shown in the left panels of Fig. 12. Subsequently, the fifth soliton, which had a small amplitude, separated from the first four solitons and lagged behind this group. The solitons in the leading group were bound by their non-monotonic tails and exhibited complex quasi-chaotic interactions (see right panels in Fig. 12). Apparently, even more complex dynamics of Ostrovsky solitons can occur when their number is large, emerging from an initial intense pulse.

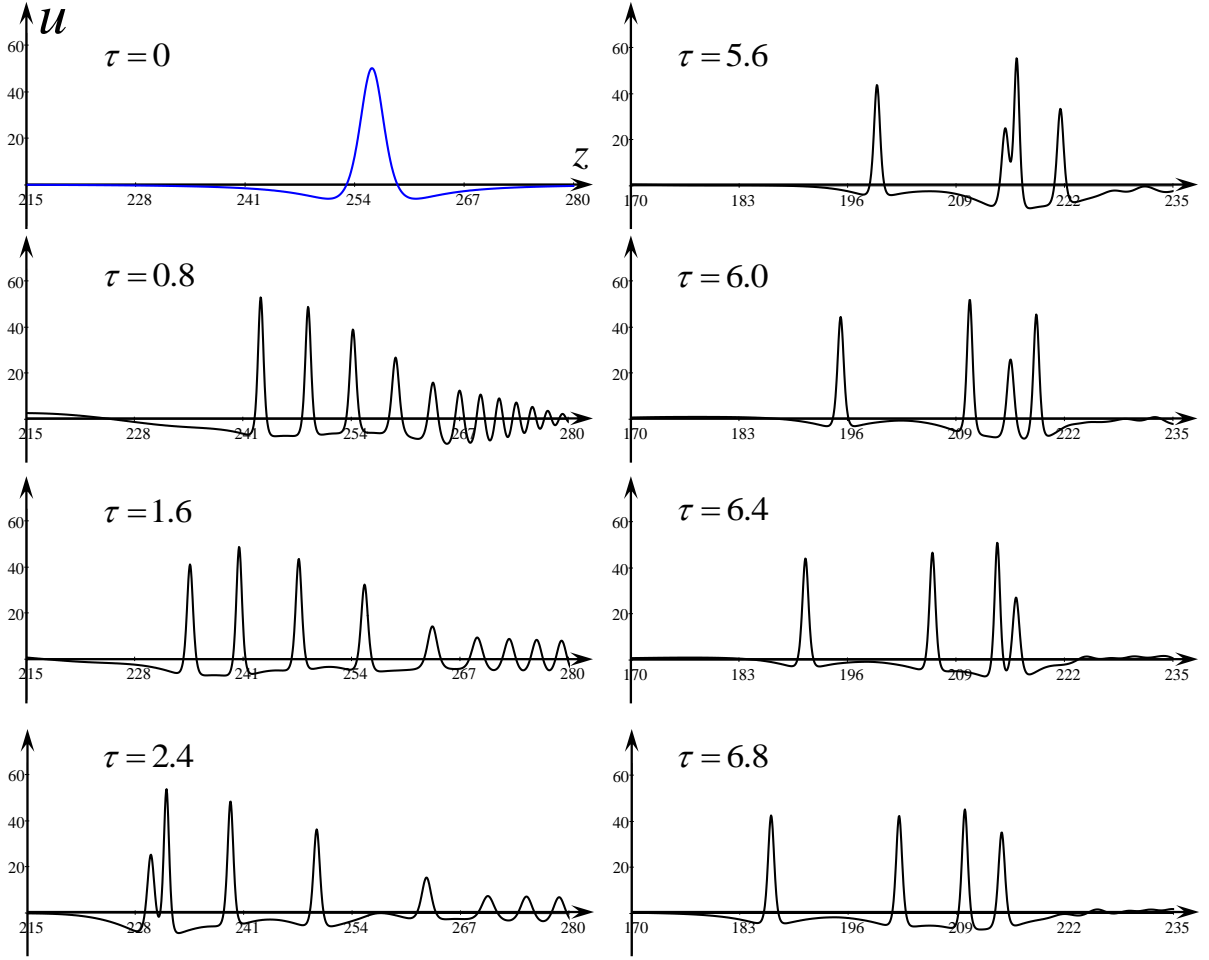

Fig. 12. Disintegration of the initial pulse (blue line in the left upper frame) into 5 solitons and a trailing ripple (left panels). The right panels demonstrate the subsequent complex dynamics of the leading four solitons bound by their non-monotonic tails.

## 4. Soliton interactions

Let us consider now the interaction of solitons of different shapes and amplitudes. As the first step, consider the interaction of big-amplitude solitons with $A_1 = 14.7$, $A_2 = 13.08$ and non-oscillating tails whose shapes are shown in the upper panel of Fig. 13 at $\tau = 0$. From Fig. 13, one can see that as the result of the interaction, solitons emit small-amplitude ripples that propagate to the left and right with different wavelengths due to the influence of two types of dispersion in the Ostrovsky equation (see the frame at $\tau = 116$). Small-scale ripples due to the Boussinesq-type dispersion propagate to the right behind solitons moving to the left, and large-scale ripples due to the Coriolis-type dispersion propagate to the left in front of solitons. The soliton amplitudes change so that after the interaction, they become $A'_1 = 16.86$, $A'_2 = 9.53$, i.e., the larger soliton becomes taller, whereas the smaller soliton becomes shorter. This is the typical scenario for inelastic soliton interaction in non-integrable systems (Krylov & Yan'kov, 1980; Zakharov *et al.*, 1988; D'yachenko *et al.*, 1989; Zakharov & Kuznetsov, 2012). As shown in the cited studies, even a small energy gain by the larger soliton resulting from a collision can be highly consequential after many interactions, for example, in a bounded confinement with multiple soliton collisions. Eventually, only one soliton ("soliton terminator") survives in such cases.

Our calculations were conducted in a closed system with periodic boundary conditions and a spatial period $L = 128$. Therefore, after the first interaction, solitons moved on a wavy background until the next interaction. After each interaction, the difference between soliton amplitudes increased until the smaller soliton completely dissolved into the background wave field, as shown in Fig. 13 at $\tau = 230$. Variations of soliton amplitudes with time are shown in Fig. 14. The first soliton interaction occurred at $\tau \approx 113$, after which the amplitude of the first soliton increased by a factor of 1.146, and the amplitude of the second soliton decreased by a factor of 0.651. The second interaction occurred at $\tau \approx 166$, after which the amplitude of the first soliton further increased by a factor of 1.055, and the amplitude of the second soliton decreased by a factor of 0.687. The third interaction occurred at $\tau \approx 202$, after which the amplitude of the first soliton further increased by a factor of 1.043, and the amplitude of the second soliton decreased to 0.2 so that the soliton became indistinguishable from the background wave field. Thus, the largest soliton in the initial condition completely terminated the second soliton in the closed system after several collisions.

Next, we considered the interaction of a big-amplitude soliton, $A_1 = 13.08$, with a small-amplitude soliton, $A_2 = 2.87$, with oscillating tails; soliton shapes are shown in the upper panel of Fig. 15 at $\tau = 0$. From Fig. 15 one can see that as the result of the interaction, solitons emit small-amplitude ripples that propagate in opposite directions as above (see the frame at $\tau = 26$). The amplitude of the larger soliton slightly increased by the factor 1.02, but the second soliton does not survive the interaction; it transforms into the ripple – see the frame at $\tau = 30$ in Fig. 15. The termination phenomenon occurred again.

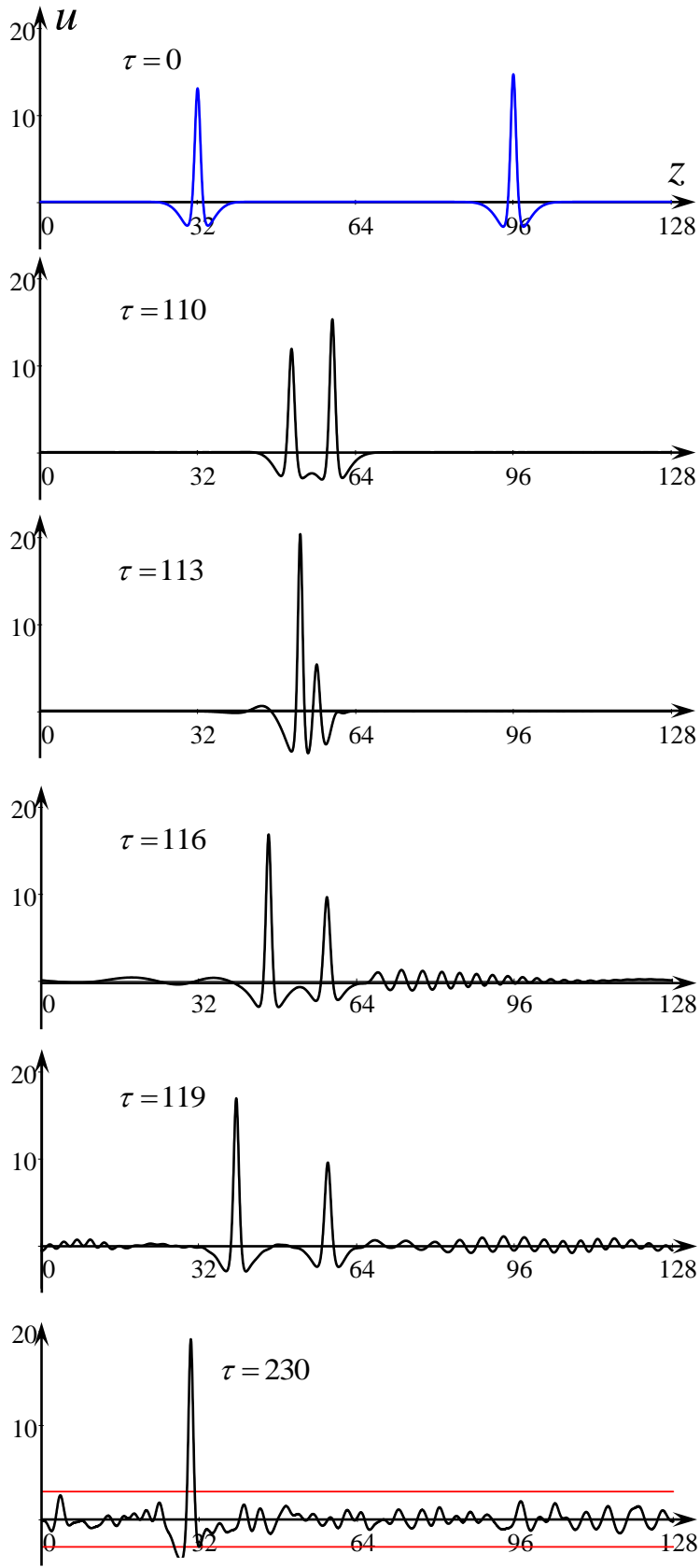

Fig. 13. Interaction of two solitons with big amplitudes, $A_1 = 14.7$ and $A_2 = 13.08$. Horizontal red lines in the bottom frame show the range of variations of the background wave field.

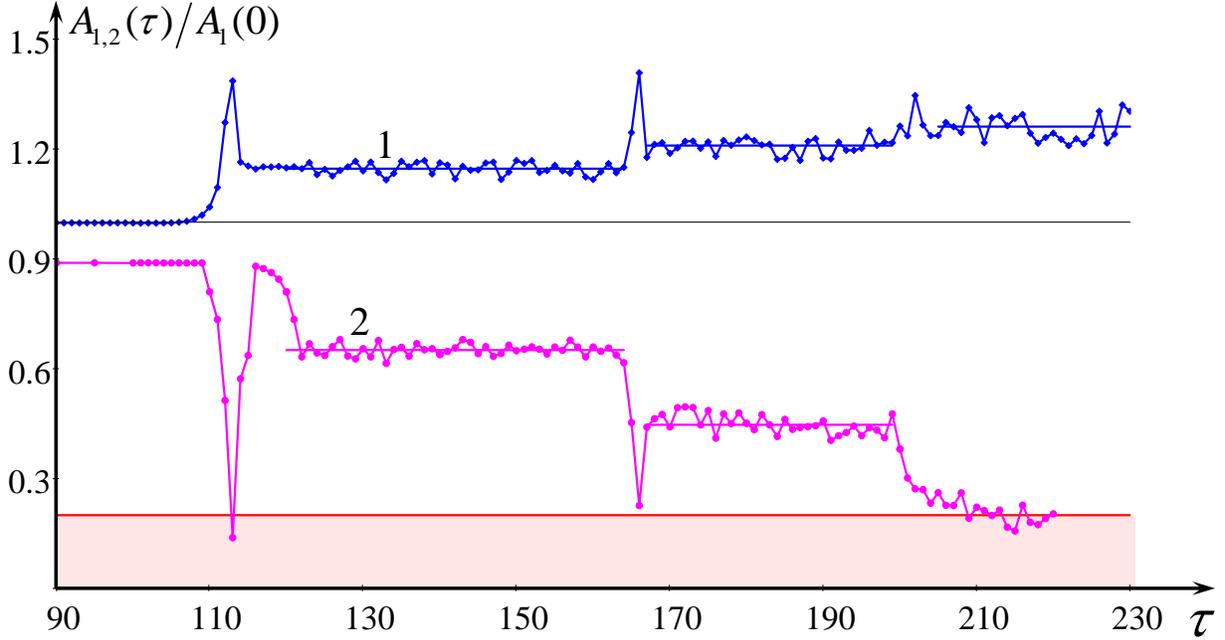

Fig. 14. Time variation of normalised soliton amplitudes with the initial values $A_1 = 14.7$ and $A_2 = 13.08$. Red horizontal line shows the upper boundary of a background wave field. Other horizontal lines show average soliton amplitudes between successive collisions.

In conclusion, we studied the interaction of a large-amplitude soliton with stable and unstable bi-solitons. As the first case, we considered the interaction of a single soliton with amplitude $A_1 = 13.08$ with a small-amplitude unstable bi-soliton with $A_2 = 6.148$ (see the upper panel in Fig. 16). These solitons moved in opposite directions: the single soliton moved to the left, and the bi-soliton moved to the right. In the course of motion, the bi-soliton split into two solitons that entered into an exchange-type interaction so that some portion of energy from the front soliton was transferred to the rear soliton. Then, the front soliton became smaller, separated from the second soliton, and started moving ahead (to the right) with a slightly higher speed than the rear one. This can be seen in the second panel at $\tau = 45$. The third panel at $\tau = 55$ shows the moment of collision between the front small-amplitude soliton and the large-amplitude single soliton moving to the left. After the collision, the small-amplitude soliton disappeared, producing ripples, and the large-amplitude soliton became a bit taller and continued moving to the left (see the panel at $\tau = 65$). The second (rear) soliton of slightly larger amplitude survived the first interaction with the large soliton (see the panel at $\tau = 80$), but subsequently it was also terminated by the large soliton after the second interaction in the closed system.

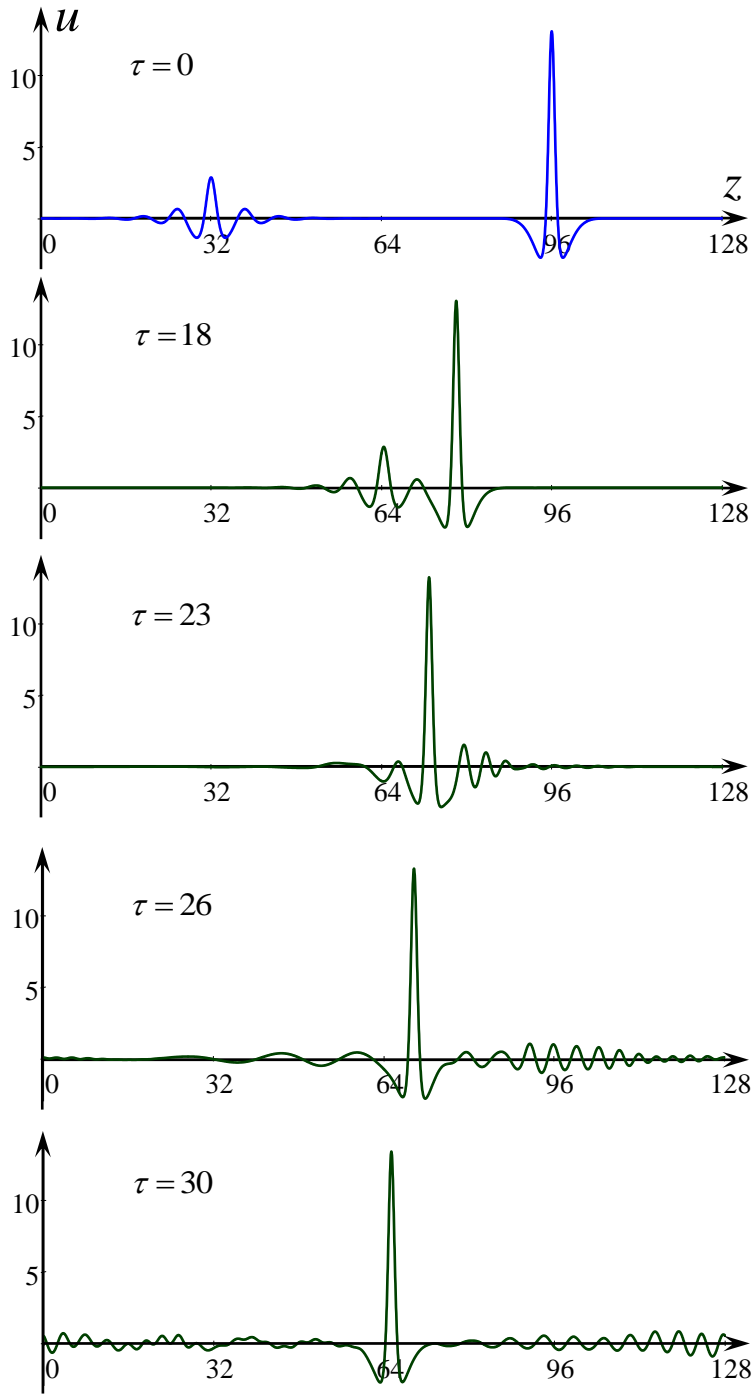

Fig. 15. Interaction of two solitons of big and small amplitudes, $A_1 = 13.08$ and $A_2 = 2.87$.

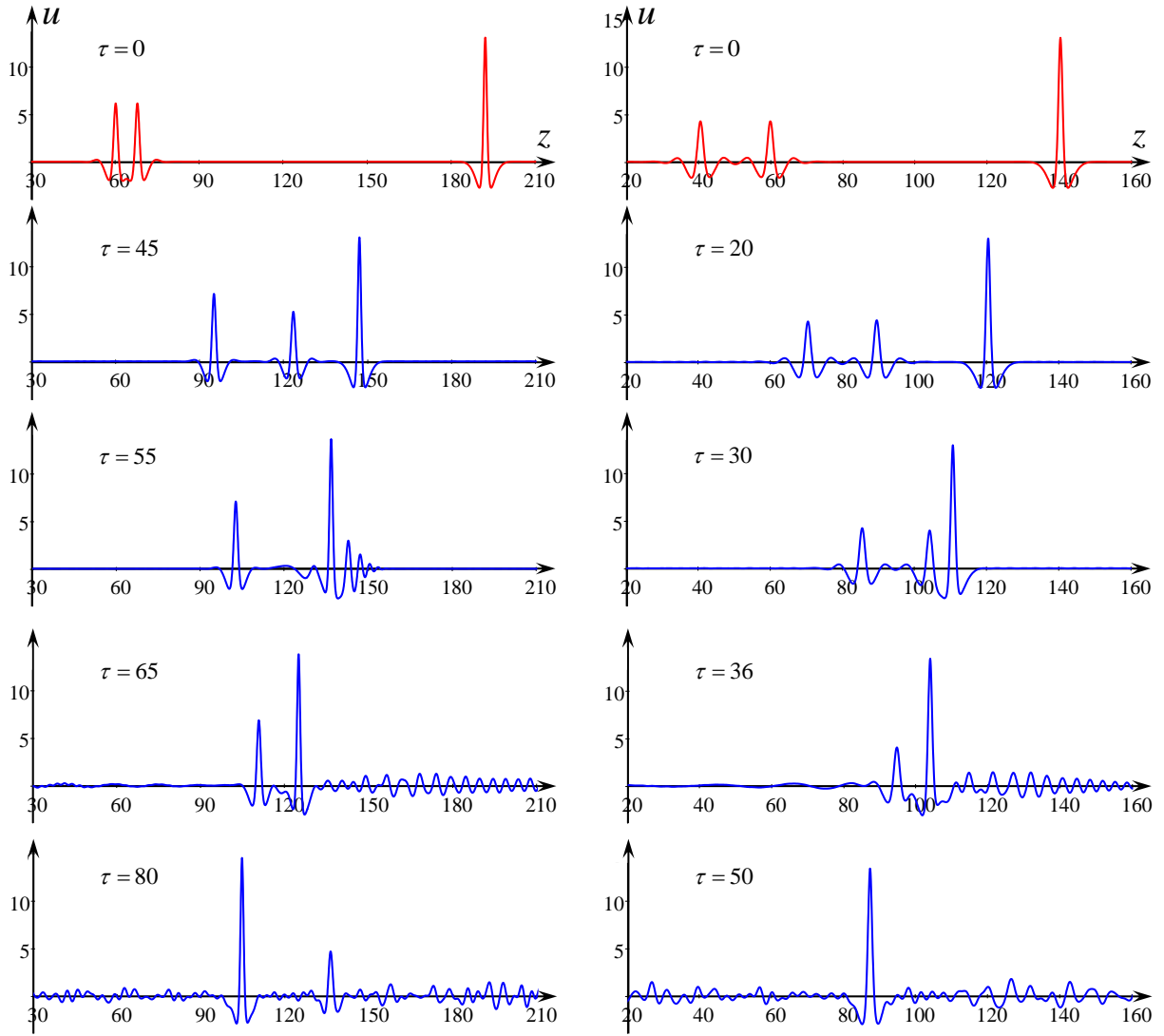

Fig. 16 (left panels). Interaction of a single soliton with the amplitude $A_1 = 13.08$ and an unstable bi-soliton with the amplitude $A_2 = 6.148$.

Fig. 17 (right panels). Interaction of a single soliton with the amplitude $A_1 = 13.08$ and a stable bi-soliton with the amplitude $A_2 = 4.278$.

Then, we considered the interaction between a single soliton with the same amplitude $A_1 = 13.08$ and a small-amplitude stable bi-soliton with $A_2 = 4.278$ (see the upper panel in Fig. 17). These solitons also moved in opposite directions, as in the previous case. In the course of motion, the bi-soliton remained stable and moved as one entity (see the panel at $\tau = 20$). At $\tau = 30$, the single and bi-solitons entered into interaction, and the front soliton in the bi-soliton pair was terminated by the single soliton, resulting in the generation of ripples (see the panel at $\tau = 36$).

After that, the single soliton also terminated the second small-amplitude soliton, taking a portion of its energy and transferring the remaining energy into ripples as shown in the panel at $\tau = 50$. Thus, we can see again that terminal interaction occurs between large- and small-amplitude solitons in the non-integrable system such that only the soliton of largest amplitude survives.

## 5. Quasi-recurrence in the Ostrovsky equation

In this section, we examine the quasi-recurrence phenomenon in the Ostrovsky equation when the initial condition is sinusoidal, $u(\xi, 0) = A \cos(\pi \xi)$ (cf. (Zabusky and Kruskal, 1965)). *Recurrence* refers to the restoration of the initial sinusoidal waveform after it disintegrates into multiple solitons that subsequently interact with each other and nearly reconstruct the original harmonic state, possibly with a phase shift. Detailed investigation of this phenomenon within the completely integrable KdV equation (see (Salupere et al., 2003)) reveals that the initial state is not perfectly restored but only approximately recovered. Nevertheless, a periodic pattern emerges in which, after sufficiently long time intervals, an almost perfect restoration occurs – a phenomenon termed *super-recurrence*.

Within the non-integrable Ostrovsky equation, where soliton interactions are inelastic, one might expect that after long-time evolution of a sinusoidal initial perturbation, the largest soliton (the "champion") would eventually survive while all other solitons are suppressed. To examine this hypothesis, we undertook a numerical study of the Ostrovsky equation (2.1) with the similar initial condition as in (Zabusky and Kruskal, 1965), $u(0, z) = A \cos[\pi(z - 1)]$ and $A = 1000$. This function produces approximately the same number of solitons as in (Zabusky and Kruskal, 1965), but in our case, they move to the left (see Fig. 18).

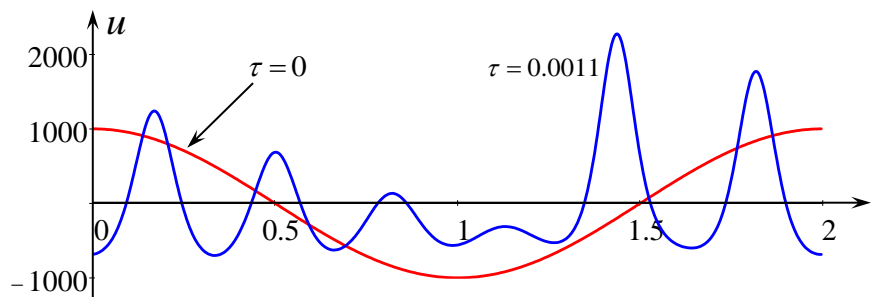

Fig. 18. Disintegration of a sinusoidal perturbation (red line) onto 6 solitons (blue line at $\tau = 0.0011$).

To reveal the recurrence from the numerical data, we examined spatial spectra of the numerical data to determine time moments when the spectrum of $u(\xi)$ is minimal. To this end, we calculated the quantity

$$S(\tau) = \frac{2}{\sqrt{N}} \sum_{i=1}^{100} \left\| |S_i(\tau)| - |S_i(0)| \right\|, \tag{5.1}$$

where $S_i(\tau)$ is a spatial spectrum of the function $u(z)$ at a fixed time, $S_i(0)$ is a spatial spectrum of $u(z)$ at $\tau = 0$, $N$ is the number of points in the series $u(z_n)$. The corresponding number of Fourier harmonics was $N/2$ but their amplitudes quickly decreased with the harmonic number $i$, therefore, we summed up only over 100 harmonics. Figure 19 shows the evolution of a spatial spectrum $S(\tau)$ with time.

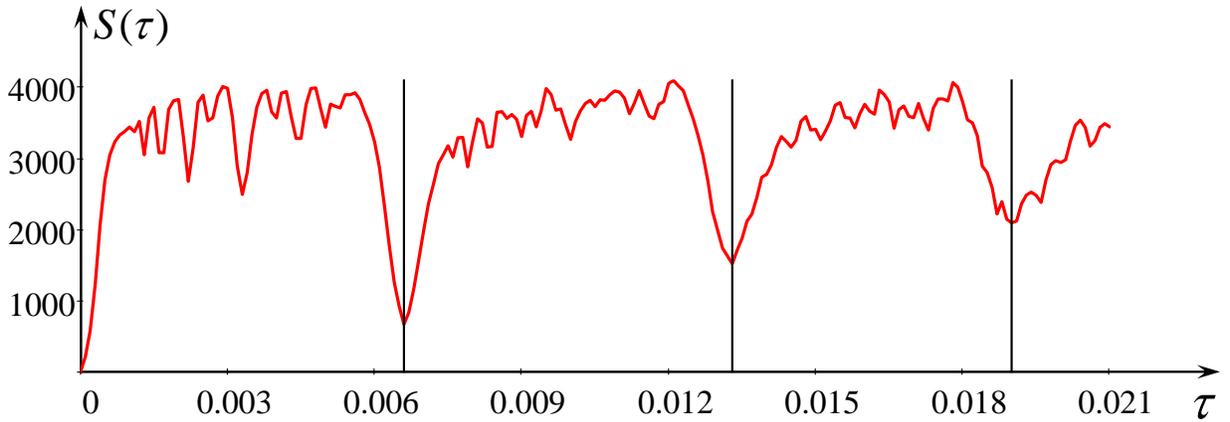

Fig. 19. Evolution of the signal spectrum as per Eq. (5.1). Vertical lines show the time moments ($\tau = 0.0066, 0.0133, 0.019$) when the spectrum width is minimal.

As shown in Fig. 19, the spectrum exhibits minimal values at certain time moments, but these minimal values do not reach zero, and increase with time. Therefore, only quasi-recurrences occur, similar to those observed in the KdV equation. Figure 20 shows the shape of the function $u(z)$ at the time moments when quasi-recurrence occurred. Possibly that after a very long computation time, a super-recurrence could emerge when the spectrum width becomes especially narrow. However, even in a much longer calculation up to $\tau = 0.7$ performed by V.A. Gordin and D.P. Milyutin using more efficient numerical code (see their Chapter in this book) with the same parameters and initial condition a super-recurrence was not observed.

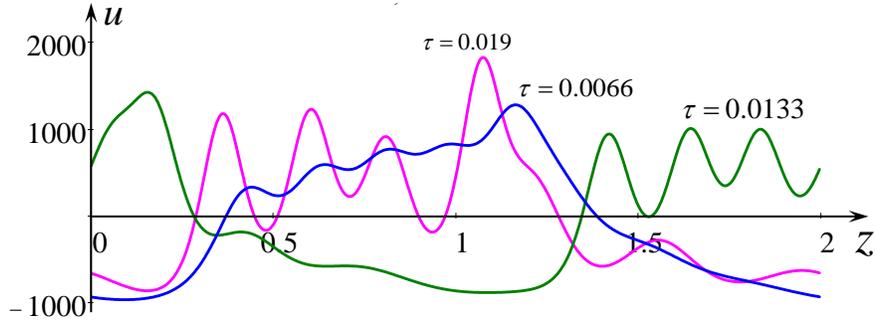

Fig. 20. Function *u*(*z*) at three time moments shown in Fig. 19 when quasi-recurrence occurs.

To our surprise, we observed neither the suppression of small solitons after interactions with large solitons nor the formation of a soliton champion. A possible explanation is that the spatial domain confining the six solitons in our computation was too short, providing insufficient room for radiation to escape after soliton collisions, causing it to contribute instead to the restoration of the initial state. This issue is worth to further study.

## 6. Conclusion

In this paper, we have studied the features of soliton formation and interaction in the non-integrable Ostrovsky equation with anomalous (positive) dispersion. We have shown that soliton interaction is inelastic, which leads to the radiation of a small-amplitude wavetrain after rhe interaction. Meanwhile, Ostrovsky solitons with zero total mass and non-monotonic profiles can arise from pulse-type initial perturbations. They can form either regular trains with solitons arranged by amplitude or irregular nonstationary confined formations of bounded interacting solitons. The conditions that determine whether pulse disintegration results in regular or irregular outcomes remain unclear.

We have also demonstrated that through multiple interactions of two solitons of different amplitudes in a closed system with periodic boundary conditions, the largest one survives with increased amplitude compared to its initial value, whereas the smaller soliton eventually disappears, transforming its energy partially into the largest soliton and partially into radiation. This leads to the formation of a soliton champion – the single soliton that survives in a closed system with periodic boundary conditions.

We have also examined the recurrence phenomenon and discovered that it resembles the analogous phenomenon in the KdV equation, but we did not observe super-recurrence, wherein the initial state is recovered with high accuracy in long-term evolution. Surprisingly, we observed neither the suppression of small solitons after interactions with large solitons nor the formation of a soliton champion. This issue warrants further study.

**Acknowledgements.** The authors are thankful to V.A. Gordin and D.P. Milyutin for the data on recurrence phenomenon obtained in the long-term calculation. R.F. acknowledges funding received from FAPEMIG (grant FCT-00208-25).

**R. Fariello**
**Departamento de Ciências da Computação, Universidade Estadual de Montes Claros, Avenida Rui Braga, sn, Vila Mauricéia, Montes Claros, 39401-089, Minas Gerais, Brazil**
**E-mail: ricardo.fariello@unimontes.br**

**M.S. Soares**
**Programa de Pós-Graduação em Modelagem Computacional e Sistemas, Universidade Estadual de Montes Claros, Avenida Rui Braga, sn, Vila Mauricéia, Montes Claros, 39401-089, Minas Gerais, Brazil**
**E-mail: magnusssoares@gmail.com**

**Y.A. Stepanyants**
**School of Mathematics, Physics and Computing, University of Southern Queensland, 487-535 West St., Toowoomba, QLD, 4350, Australia**
**E-mail: Yury.Stepanyants@unisq.edu.au**